\documentclass[trackchanges,twocolumn,tighten]{aastex63}

\newcommand{\ltsima}{$\; \buildrel < \over \sim \;$}
\newcommand{\simlt}{\lower.5ex\hbox{\ltsima}}

\def\arcdeg{\hbox{$^\circ$}}
\def\arcmin{\hbox{$^\prime$}}
\def\arcsec{\hbox{$^{\prime\prime}$}}

\usepackage{hyperref}

\usepackage{amsmath}
\usepackage{graphicx}
\turnoffedit

\shorttitle{An X-ray view of NGC~1266}
\shortauthors{Lopez et al.}

\begin{document}

\title{The Spatially Resolved Hot Gas Properties of NGC~1266's AGN-Driven Outflow}

\correspondingauthor{Sebastian Lopez}
\email{lopez.764@osu.edu}

\author[0000-0002-2644-0077]{Sebastian Lopez}
\affiliation{Department of Astronomy, The Ohio State University, 140 W. 18th Ave., Columbus, OH 43210, USA}
\affil{Center for Cosmology and AstroParticle Physics, The Ohio State University, 191 W. Woodruff Ave., Columbus, OH 43210, USA}

\author[0000-0002-1790-3148]{Laura A. Lopez}
\affil{Department of Astronomy, The Ohio State University, 140 W. 18th Ave., Columbus, OH 43210, USA}
\affil{Center for Cosmology and AstroParticle Physics, The Ohio State University, 191 W. Woodruff Ave., Columbus, OH 43210, USA}

\author[0000-0002-3249-8224]{Lauranne Lanz}
\affil{Department of Physics, The College of New Jersey, 2000 Pennington Road, Ewing, NJ 08628, USA}

\author[0000-0003-3191-90396]{Justin A. Otter}
\affil{William H. Miller III Department of Physics and Astronomy, Johns Hopkins University, Baltimore, MD 21218, USA}

\author[0000-0002-4261-2326]{Katherine Alatalo}
\affil{William H. Miller III Department of Physics and Astronomy, Johns Hopkins University, Baltimore, MD 21218, USA}
\affil{Space Telescope Science Institute, 3700 San Martin Drive, Baltimore, MD 21218, USA}

\begin{abstract}

Galactic winds play a critical role in galaxy evolution, yet their structure and driving mechanisms remain poorly understood, especially in low-luminosity AGN (LLAGN) systems. NGC~1266 hosts one such LLAGN, embedded in a massive molecular gas reservoir that is not forming stars, likely due to AGN feedback. We analyze deep archival \textit{Chandra} data to constrain the properties of its hot gas and compare them to other wind systems. We find temperatures of \edit1{$0.25-1.8$~keV} and notably high electron densities of $0.33-4.2$~cm$^{-3}$, suggesting significant mass loading, further supported by charge exchange emission in the southern lobe, one of the few AGN systems where it has been detected. We measure pressures and thermal energies of $10^6-10^8$~K cm$^{-3}$ and $10^{54}-10^{56}$~erg, exceeding the minimum energy needed for the radio jet to power the outflow and implying the hot phase comprises a large fraction of the energy budget. Archival MUSE data reveal a cavity-like feature in the southern outflow, potentially associated with the far side of the outflow cone. At the maximum outflow extent, the warm and hot phases appear to be in pressure equilibrium. Coupled with short cooling timescales of $\sim$1~Myr, comparable to the advection time, this suggests the outflow is undergoing radiative cooling and may have stalled. Finally, we compare NGC~1266 to other local AGN and starburst galaxies, finding NGC~1266 to be the densest wind in the sample.

\end{abstract}


\section{Introduction} \label{sec:intro}

Galactic feedback is necessary to explain various observed properties in the Universe such as the absence of ultra-massive galaxies in cosmological simulations \citep{Oppenheimer2010}, the stellar-mass metallicity relationship \citep{Tremonti2004}, and the enrichment of the circum- and inter-galactic medium surrounding galaxies \citep{Oppenheimer2008}. This feedback, in the form of gas outflows, or winds, and can be driven either through a starburst event like in the cases of M82 and NGC~253 \citep{Lopez2020,Lopez2023}, or via active galactic nuclei (AGN) feedback. Regardless of origin, galactic winds are multiphase phenomenon requiring a multi-wavelength approach (X-ray, optical, radio, infrared, and ultraviolet studies) to understand the various physics occurring, such as the ejection of metal-rich gas from the \edit1{galaxy's interstellar medium (ISM)}, the quenching of star formation, and stellar abundance trends in the host galaxy's \edit1{stellar} disk \citep{Veilleux2005,Veilleux2020, Thompson2024}.

\begin{table*}
\centering
\caption{Basic properties of NGC\,1266.}
\label{tab:ngc1266_properties}
\begin{tabular}{llr}
\hline\hline
Property & Value & Reference \\
\hline
R.A.\ (ICRS)             & $03^{\rm h}16^{\rm m}00\fs8$           & \cite{Gaia2020} \\
Dec.\ (ICRS)             & $-02\degr25\arcmin38\arcsec$            & \cite{Gaia2020} \\
Morphology       & S0                                      & \cite{HyperLeda} \\
Distance                 & $29.9$\,Mpc      & \cite{Cappellari2011} \\
Inclination             & 58.2$\degr$         & \cite{HyperLeda} \\
$m_B$ & $13.8$\,mag                             & \cite{HyperLeda} \\
$\mathrm{D_{25}}$       & $1.3\arcmin$          & \cite{HyperLeda} \\
Stellar mass    & $6.5\times10^{9}\,M_\odot$ &\cite{Leroy2019} \\
Black hole mass  & $1.7\times10^{6}$\,$M_\odot$  & \cite{Alatalo2015} \\
Molecular Outflow Mass & $2.8\times10^8\;M_\odot$ & \cite{Otter2026} \\
Hot Gas Outflow Mass\tablenotemark{a} & $1.35\times10^8\;\sqrt{f}\;M_\odot$ & This Paper \\
$L_{\rm FIR}$ & $1.3\times10^{10}$\,$L_\odot$   & \cite{Alatalo2015} \\
$L_{\rm X}$ & $1.5\times10^{7}$\,$L_\odot$   & This Paper \\
Current SFR              & $\sim 0.7$\,$M_\odot$\,yr$^{-1}$ & \cite{Otter2024} \\
\hline
\end{tabular}
\tablenotetext{a}{$f$ is the filling factor of the hot gas that we assume is unity.}
\end{table*}

In this paper, we focus on an extreme example of AGN feedback launching tremendous outflows that stifle star formation in the galaxy NGC~1266. It is a nearby (29.9~Mpc; \citealt{Cappellari2011}), early-type galaxy (ETG) that has a massive CO outflow \citep{Alatalo11} discovered in the $\mathrm{ATLAS^{3D}}$ survey \citep{Cappellari2011} \edit1{and a summary of its general properties can be found in Table~\ref{tab:ngc1266_properties}}. The outflow has been studied several times in the optical and mm wavelengths: \cite{Alatalo11} first reported the discovery of the massive molecular gas reservoir $\sim\;10^9\;M_{\odot}$ as well as an outflow driving 13 $\mathrm{M_\odot/yr}$ from the \edit1{galaxy's ISM}. \cite{Alatalo2015} later amended the estimate to \edit1{110 $\mathrm{M_\odot/yr}$, assuming the outflowing gas is optically thick; the most recent ALMA observations of \citealt{Otter2026} revise this downward to ${<}85~\mathrm{M_\odot/yr}$ under a multiphase, clumpy model in which the $^{12}$CO traces a diffuse optically thin component while dense clumps are traced by HCN(1--0)}. \edit1{AGN-driven molecular outflows are most commonly found in luminous infrared galaxies and quasar hosts \citep{Veilleux2020, Fluetsch2019}, making NGC~1266's outflow exceptional given its modest stellar mass and lack of merger signatures \citep{Alatalo11}.} The dearth of star formation is an indication that the outflow cannot be stellar feedback driven. \cite{Alatalo2014}, using SAURON as well as UVOT, confirmed the galaxy is a post-starburst system with the youngest stellar populations confined to the central 1 kpc and current star formation constrained to the inner few hundred parsecs, indicating no large-scale star formation can drive an outflow.

\begin{figure*}
    \centering
    \includegraphics[width=\textwidth]{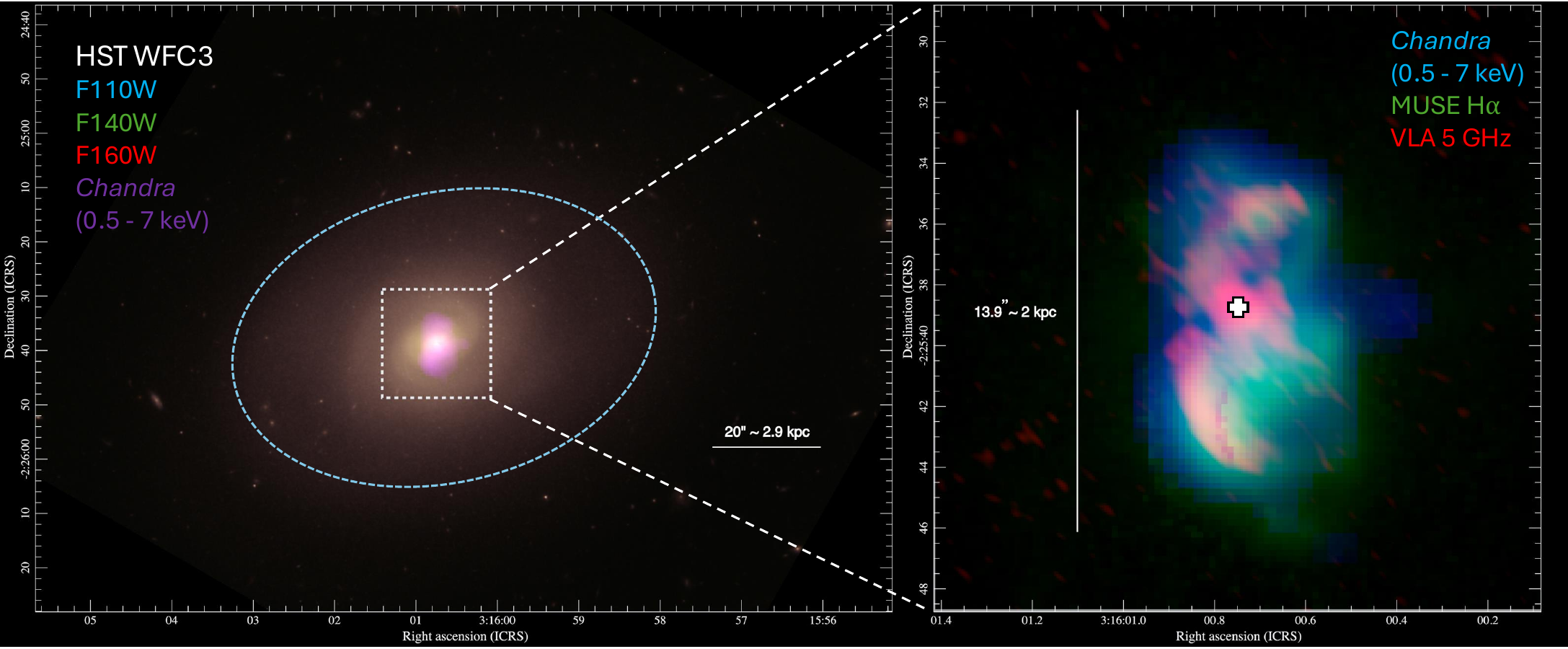}
    \caption{\textit{Left}: Four color image of NGC~1266. The HST data show the disk in the F110W (blue), F140W (green), and F160W (red) filters while the \textit{Chandra} X-ray data (violet) show the hot outflow. \edit1{The light blue ellipse is the $D_{25}$ isophotal diameter of 1.3\arcmin{} from \cite{HyperLeda}.} The white box is 20\arcsec{} by 20\arcsec{} and is the area of focus for this analysis. \textit{Right}: Zoom-in of the white box in the left panel to highlight the multiwavelength nature of the outflow. In blue is the broad-band ($0.5-7$~keV) \textit{Chandra} data, green is the optical H$\alpha$ data from MUSE, and red is the 5~GHz radio continuum data. The white cross marks the location of the AGN \citep{Nyland2013}. At the distance of NGC~1266 (29~Mpc; \citealt{Cappellari2011}), 1\arcmin{} is about 8.6~kpc.}
    \label{fig:3color}
\end{figure*}

The outflow in NGC~1266 must therefore be driven by an AGN. \cite{Nyland2013} used VLA and VLBA to study the radio emission of the outflow and found a high brightness temperature radio core at the center of the galaxy indicative of an AGN. \edit1{\cite{Nyland2013} also found the radio emission to be consistent with other low-luminosity AGNs (LLAGNs), showing that even low-powered AGNs can drive strong outflows.}  \cite{Davis2012} used SAURON and GMOS IFU data to study the outflow, finding it emits in optical lines, and constrained the outflow kinematics to velocities up to 900 km~s$^{-1}$. They also found extended low ionization emission that they attributed to shocks of the radio jet with the ambient ISM. \cite{Otter2024} studied the inner 500~pc of NGC~1266 using Gemini-NIFS and also found shocks by probing several H$_2$ rovibrational emission lines. These investigations, along with X-ray observations shown in \cite{Alatalo2015}, demonstrate the complex, multiphase nature of NGC~1266. In Figure~\ref{fig:3color} we show the various multiwavelength data of NGC~1266 with the HST near-infrared filters (F110W, F140W, F160W; doi:\dataset[10.17909/wr5m-7h21]{https://doi.org/10.17909/wr5m-7h21}) and \textit{Chandra} X-ray data on the left, and the zoom in of the outflow on the right showing the X-ray, MUSE H$\alpha$, and VLA 5 GHz radio.  

Since \cite{Alatalo2015}, new \textit{Chandra} X-ray observations have been taken of NGC~1266 that, in conjunction with archival data, amount to 150~ks of data. These data now allow for a detailed analysis of the hot phase of NGC~1266's wind and its relation to the colder phases. In this paper we conduct a spatially-resolved analysis of the X-ray spectra to constrain the temperature and volume density distributions and to make estimates of energetics of the system, comparing them to the cooler wind phases. We also take advantage of archival MUSE data to provide further insight into the ionization conditions of the wind at higher resolution than that achieved in \cite{Davis2012}. 

The paper is structured as follows. We describe our methodology in Section~\ref{sec:methods}, and we present the hot gas properties in Section~\ref{sec:results}. We discuss the implications and compare the hot gas to other phases in Section~\ref{sec:disc}. We summarize our findings in Section~\ref{sec:conclusion}.

\section{Data \& Methods} \label{sec:methods}

\begin{deluxetable}{lrcc}
\tablecolumns{4}
\tablewidth{0pt} \tablecaption{{\it Chandra} Observations Used \label{table:data}} 
\tablehead{\colhead{ObsID} & \colhead{Exposure} & \colhead{UT Start Date} &  \colhead{Instrument}}
\startdata
11578  & 30~ks & 2009-09-20 & ACIS-S \\
19498  & 80~ks & 2016-10-10 & ACIS-S \\
19896  & 40~ks & 2016-10-13  & ACIS-S
\enddata
\end{deluxetable}

\subsection{Chandra X-ray Data}

\begin{figure*}
    \centering
    \includegraphics[width=\textwidth]{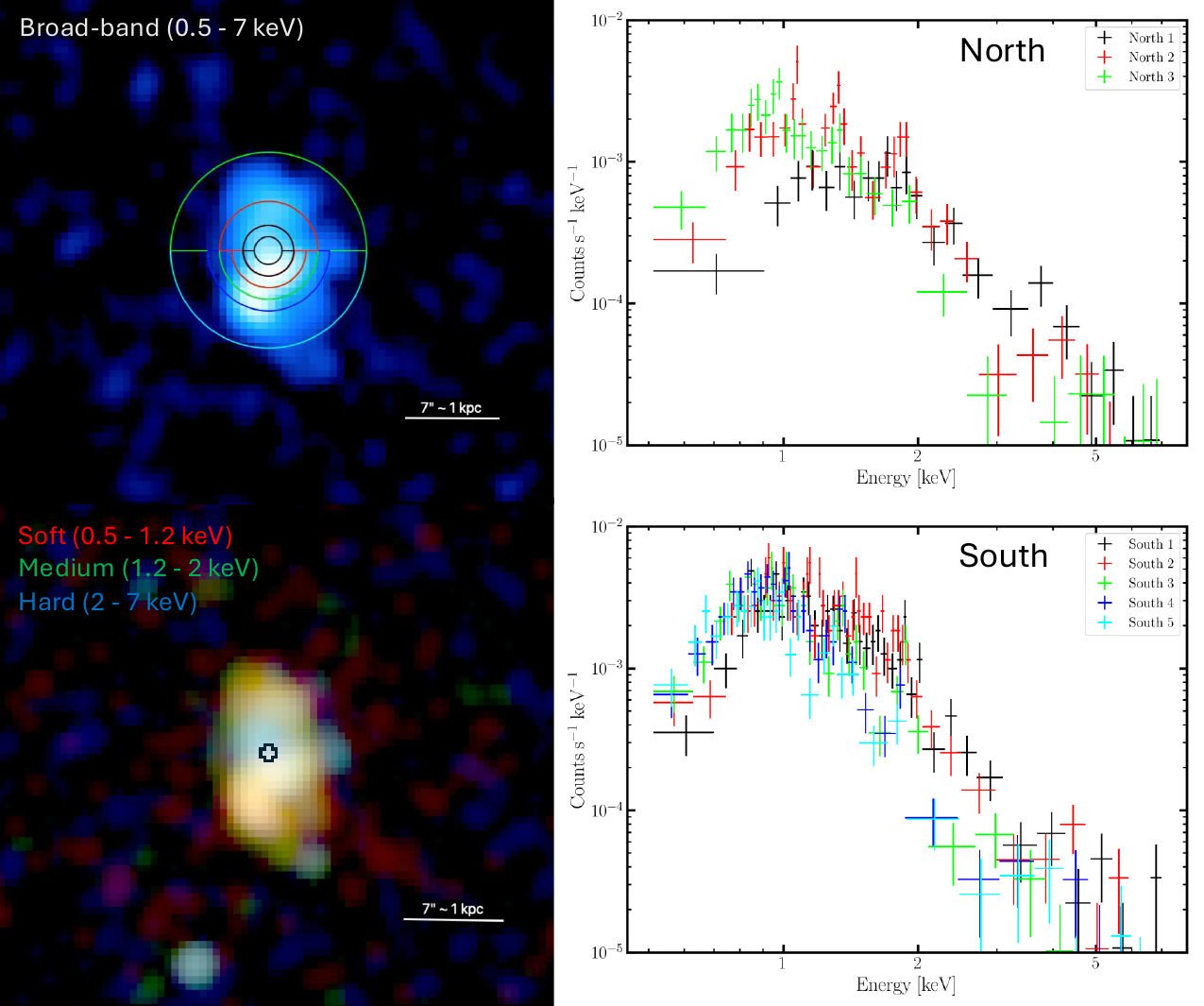}
    \caption{\edit1{\textit{Top Left:} Broad-band ($0.5-7.0$~keV) X-ray image of NGC~1266 overlaid with the regions used for spectral extraction. These regions span radii between one and seven arcseconds ($\approx265$--$1013$~pc) and are color-coded to match the spectra shown in the right panels. At a distance of 29~Mpc, $7\arcsec \approx 1$~kpc, as indicated by the scale bar. In these images, north is up and east is to the left. \textit{Bottom Left:} Exposure-corrected, three-color image of NGC~1266, where red corresponds to soft ($0.5$--$1.2$~keV), green to medium ($1.2$--$2.0$~keV), and blue to hard ($2.0$--$7.0$~keV) X-ray emission. Yellow highlights soft-to-medium emission tracing the diffuse gas, while blue traces the nuclear emission as well as point sources. The AGN is marked by a cross in the three-color image. \textit{Top Right:} The extracted and combined spectra for each of the northern annuli, colored according to their corresponding regions shown in the top left panel. \textit{Bottom Right:} Same as the top right panel, but for the southern regions.}}
    \label{fig:xray_image}
\end{figure*}

NGC~1266 was observed three times with {\it Chandra} in 2009 and 2016 with ACIS-S, as detailed in Table~\ref{table:data} (doi:\dataset[10.25574/cdc.458]{https://doi.org/10.25574/cdc.458}). Data were downloaded from the \textit{Chandra} archive and reduced using the Chandra Interactive Analysis of Observations {\sc ciao} version 4.15 \citep{CIAO2006}. The observations were combined using the CIAO task \textit{merge\_obs} and then \textit{wavdetect} was run to identify point sources that were removed using the \textit{dmfilth} task. The resulting broad-band (0.5-7~keV) X-ray image is shown in Figure~\ref{fig:xray_image} along with a three-color image showing the different X-ray bandpasses of soft ($0.5-1.2$~keV), medium ($1.2-2.0$~keV), and hard ($2.0-7.0$~keV) X-rays.

\begin{figure*}
    \centering
    \includegraphics[width=\textwidth]{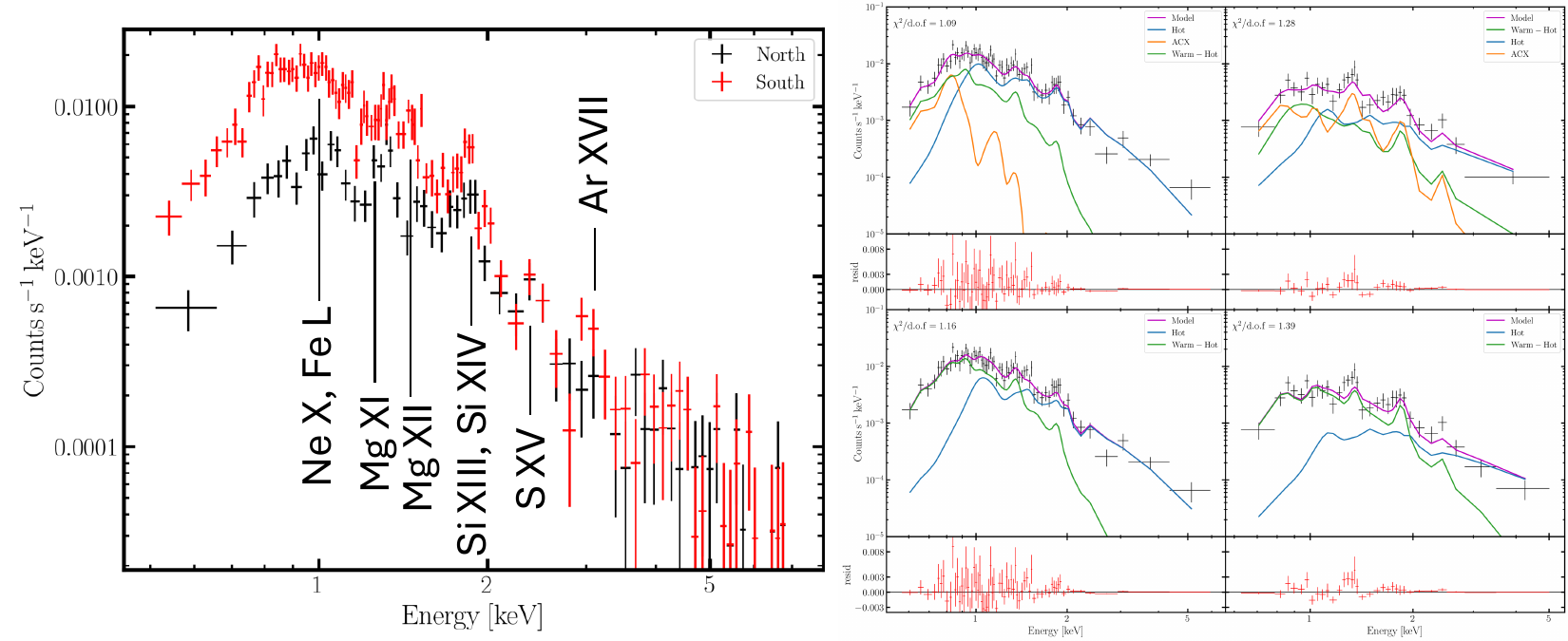}
    \caption{\edit1{ \textit{Left:} Combined spectra from all three observations extracted from the northern regions and the southern regions (i.e. encompassing all the three north and five south regions in Figure~\ref{fig:xray_image}, respectively). We label the various emission lines evident in the spectra. Spectra extracted from the composite southern region from observation 19498. \textit{Top Middle}: The CX emission is in orange (ACX), the warm-hot thermal plasma component (APEC) is in green, and the hot thermal plasma component is in blue. The data points are in black crosses, and the residuals to the best-fit model are given below. We find that in the south, a significant portion of the broad-band X-ray emission ($\sim38$\%) is dominated by CX emission. \textit{Bottom Middle}: Same as the top middle panel but the model does not include CX. Without CX, the model has larger residuals at soft X-ray energies and yields a statistically worse fit. \textit{Top Right and Bottom Right:} Same as middle panels but for the composite northern region. We do not find any improvement in the fitting when using the ACX model.}}
    \label{fig:cx}
\end{figure*}

To derive the hot phase properties of the outflow, we extracted the spectra from several regions along the minor axis of the galaxy. These regions are shown by the arcs in the \edit1{top-left} panel of Figure~\ref{fig:xray_image}. Each region was adjusted to have at least 100 net counts to properly constrain the temperature and volume density. The center of the galaxy hosting the AGN is excluded from the data as our analysis focuses on the wind (the full AGN analysis will be in Lanz et al. in prep).  Background regions were defined and subtracted from each source region. The background regions were a total area of 1.2\arcmin{} squared and located $\geq6.4$~kpc (0.74\arcmin{}) away from the galactic wind. 

The spectra were extracted using the CIAO task \textit{specextract}, and we conducted the modeling with XSPEC. We fit the spectra from each observation simultaneously. For each of the annuli shown in right panel of Figure~\ref{fig:xray_image}, we modeled the emission as an absorbed, thermal plasma in collisional ionization equilibrium (CIE). In XSPEC, these model components were \textsc{CONSTANT*TBABS*TBABS*(APEC)}. The constant is to account for emission variations between each observation. The two absorption components account for both the galactic absorption from the Milky Way, which we froze to $N_{\rm{H}}^{\rm{MW}}=6.7\times10^{20}\;\rm{cm^{-2}}$ \citep{HI4PI} and for the intrinsic absorption of NGC~1266 ($N_{\rm{H}}^{\rm{1266}}$) which we let vary freely. Both \textsc{TBABS} and \textsc{APEC} components have their abundances set to solar, consistent with studies of post-starburst galaxies \citep{Leung2024}. Our abundances are set to those of \cite{Wilms2000}, and we use the photoionization cross sections from \cite{Verner1996}. We also found the annuli regions did not benefit from an added power-law continuum to the fit where the reduced $\chi^2$ did not improve significantly statistically. \edit1{We show the full the results of the annuli regions in Table~\ref{tab:ngc1266_gradients}.}

We also tested for the presence of charge-exchange (CX) that has been found to make a significant portion of the total broad-band X-ray emission in various galactic wind hosts \citep{Lopez2020,Lopez2023,Porraz2024}. CX is the process in which an ion strips an electron from a neutral atom and reflects the interaction between hot ionized and cool neutral media. We tested a composite northern region (with all the northern annuli included) and a composite southern region for CX presence using the model component \textsc{ACX} from \cite{ACX}. With the added signal, we also found our regions benefited from a second \textsc{APEC} component. We found that the inclusion of the \textsc{ACX} component statistically improved the spectral fits of the southern region but not the north based on F-tests. The statistical improvement of the fits along with the full models are shown in Table~\ref{table:specfitscx}.

\edit1{In order to estimate the uncertainties on the best-fit values of 
Tables~\ref{table:specfitscx} and \ref{tab:ngc1266_gradients}, we employ 
the \textit{chain} command in XSPEC with the Goodman-Weare ensemble sampler 
to sample the posterior parameter space via Markov chain Monte Carlo (MCMC). 
We ran 40 walkers for $3\times10^5$ iterations and discarded the first $5\times10^4$ iterations as a burn-in to ensure the chain had converged. The reported 
uncertainties are 90\% confidence intervals.}

\begin{deluxetable*}{llccccccc}
\tablecolumns{8}
\tablewidth{0pt} 
\tablecaption{Spectral Fits for Total North and South Regions \label{table:specfitscx}} 
\tablehead{
\colhead{Region} & \colhead{Model\tablenotemark{a}} & \colhead{$N_{\mathrm{H}}$} & \colhead{$kT_1$} & \colhead{$kT_2$} & \colhead{norm$_1$} & \colhead{norm$_2$} & \colhead{$\chi^2$/d.o.f} & \colhead{F-test $p$\tablenotemark{b}} \\
\colhead{} & \colhead{} & \colhead{$(10^{22}\,\rm{cm^{-2}})$} & \colhead{$(\rm{keV})$} & \colhead{$(\rm{keV})$} & \colhead{$(\rm{cm^{-5}})$} & \colhead{$(\rm{cm^{-5}})$} & \colhead{} & \colhead{}
}
\startdata
North & APEC+APEC        & $0.71^{+0.16}_{-0.10}$     & $0.49_{-0.18}^{+0.10}$  & $3.2_{-1.2}^{+0.68}$  & $5.9_{-2.0}^{+11}\times10^{-3}$ & $1.3_{-0.30}^{+0.71}\times10^{-3}$ & 84/60  \\
North & APEC+APEC+ACX\tablenotemark{c}    & $0.53^{+0.63}_{-0.07}$ & $0.71_{-0.55}^{+0.10}$ & $2.4_{-1.4}^{+3.5}$   & $2.9_{-1.6}^{+137}\times10^{-3}$ & $3.5_{-1.4}^{+7.6}\times10^{-3}$ & 79/59 & 0.06 \\ 
South & APEC+APEC        & $0.77^{+0.13}_{-0.09}$  & $0.25^{+0.05}_{-0.03}$ & $1.3^{+0.52}_{-0.16}$ & $1.4_{-0.53}^{+1.1}\times10^{-3}$ & $1.13_{-0.32}^{+0.20}\times10^{-4}$ & 165/142  \\
South & APEC+APEC+ACX    & $0.85_{-0.08}^{+0.21}$ & $0.22^{+0.02}_{-0.05}$ & $1.1^{+0.09}_{-0.13}$ & $1.1_{-0.48}^{+3.2}\times10^{-3}$ & $1.4_{-0.19}^{+0.44}\times10^{-4}$ & 153/141 & $9.8\times10^{-4}$ \\
\enddata
\tablenotetext{a}{All models have a constant (CONST) component, two absorption components (TBABS), and metal abundances (Z) are set to solar \citep{Wilms2000}.}
\tablenotetext{b}{The significance of the F-test $p$ value is calculated as $1-p$, where $3\sigma$ confidence is 0.997.}
\tablenotetext{c}{The ACX component has the parameters $kT$, He fraction, abundance (Z), solar wind charge exchange (swcx), model, and norm. We tie the $kT$ and Z of the ACX component to the cooler thermal plasma as that is the component expected to interact with the neutral material. The He fraction is set to the default cosmic value of 0.090909. The swcx is set to 0 as we do not expect the solar wind to contribute to the data, and model is set to 8 which is the default that  sets how the code assumes electrons from the neutral participant in the CX land on the ion \citep{ACX}. The ACX norm values are $\mathrm{norm_{south}}=8.9\times10^{-6}\;\mathrm{cm^{-5}}$ and $\mathrm{norm_{north}}=3.2\times10^{-5}\;\mathrm{cm^{-5}}$}
\end{deluxetable*}

\subsection{MUSE Optical Data}

The data from the VLT's Multi Unit Spectroscopic Explorer (MUSE) IFU instrument was acquired from the ESO Science Archive, and thus the raw data was processed by the standard MUSE pipeline \citep{MUSEPipe2020}. The data covers a 1\arcmin{} by 1\arcmin{} area on the sky centered on NGC~1266 and was observed as part of program ID 0102.B-0617 (PI: A. Fl\"{u}tsch) on 24 January 2019 for 600~s. The data cubes covers $\lambda=475-935$~nm at $R=3026$, allowing for high resolution images (average seeing of 0.77\arcsec{}) of various lines that trace electron density and temperature that will complement the X-ray data. A full optical line and kinematic analysis will be presented in future work (Otter et al. 2026 in prep.); for now we focus on the H$\alpha$, H$\beta$, \ion{N}{2}, and \ion{S}{2} doublet to gain insight into the ionization conditions, outflow geometry, and warm phase temperature and density. 

In order to measure accurate line intensities, we subtracted the stellar continuum using the PHANGS-MUSE pipeline detailed in \cite{Emsellem2022}. Briefly, the pipeline runs a penalized pixel-fitting (PPXF) method from \cite{Cappellari2017} on Voronoi-binned data using the stellar population templates from the E-MILES library \cite{EMILES}. The algorithm is run multiple times to constrain the most optimal stellar kinematics, age, and metallicity to accurately model the stellar continuum being produced and allowing for it be subtracted from the cube to isolate the line emission. The continuum subtracted cube is what we use for the remainder of the analysis. 

\section{Results} \label{sec:results}

\subsection{The Hot Outflow Phase}

\begin{figure*}
    \centering
    \includegraphics[width=\textwidth]{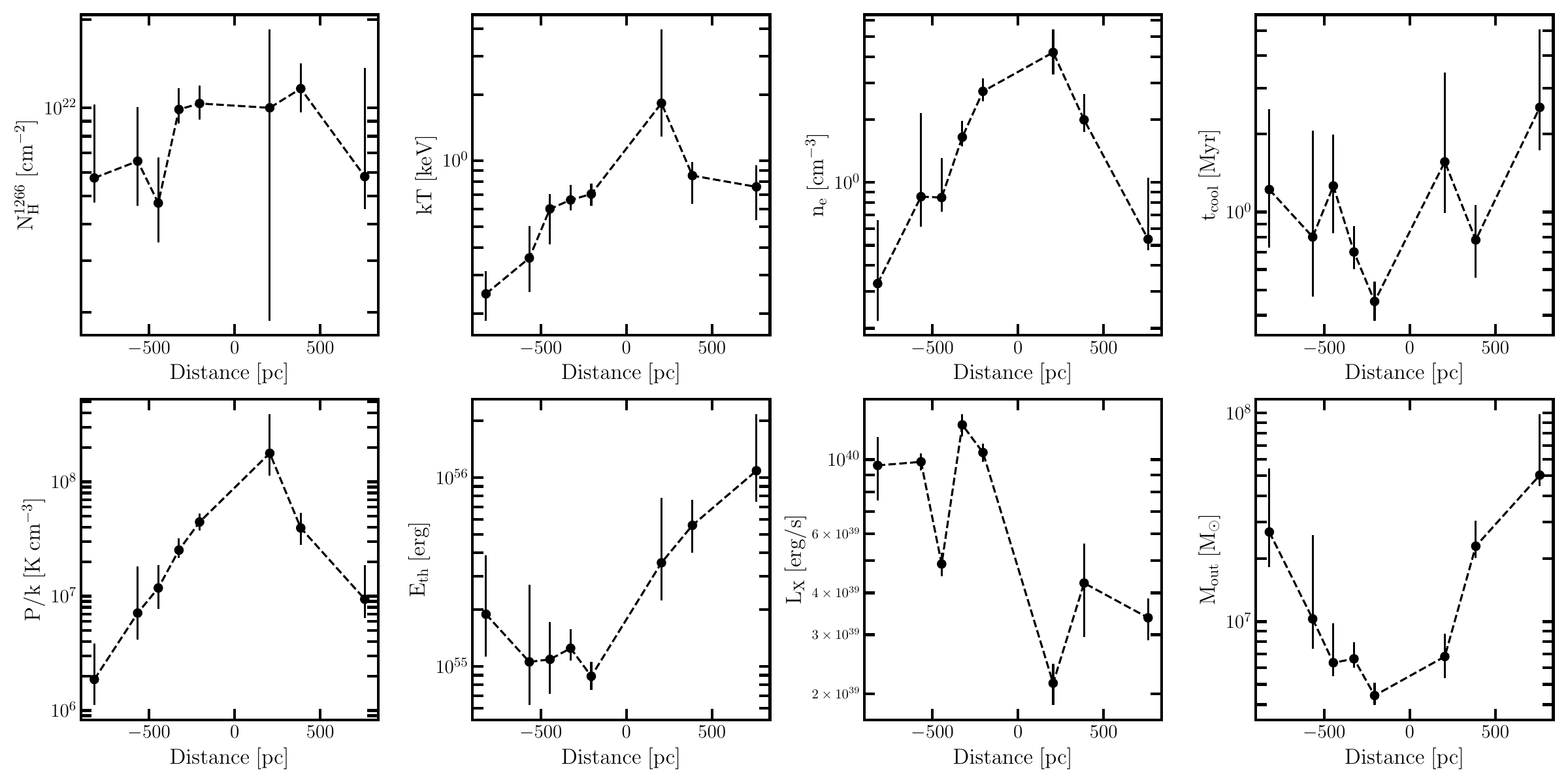}
    \caption{Gradients of measured quantities of NGC~1266's outflow. \textit{Top Row}: From the left to right, we plot the best-fit column density $N_{\rm H}^{1266}$ and temperature $kT$, derived electron number density $n_{\rm e}$ assuming spherical shell geometries for each region, and the hot gas cooling time $t_{\rm cool}$. \textit{Bottom Row}: From left to right, we plot the derived hot gas pressures $P/k$, thermal energies $E_{\rm th}$, X-ray luminosities $L_{\rm X}$, and outflow gas mass $\mathrm{M_{out}}$. Positive distances are north of the center, and negative distances are south.}
    \label{fig:gradients}    
\end{figure*}

In Figure~\ref{fig:cx}, we show the extracted spectra for the combined northern and southern regions of NGC~1266. Evident in the spectra are emission lines of Ne, Mg, Si, S, and Ar. We do not find the Fe K$\alpha$ line that is characteristically present in AGN systems since our regions did not include the central nucleus. We also notice the normalization in the $<$1.5~keV energy range is lower for the north than for the south despite comparable signal at $>$1.5~keV energies. This is an indicator of harder X-rays in the north and is evident in the three-color X-ray image in Figure~\ref{fig:xray_image}. \edit1{The lower normalization may be due to increased absorption of soft X-rays indicative of increased gas column density in the north (see Figure~\ref{fig:gradients} and Table~\ref{tab:ngc1266_gradients}).}

As described in the previous section, we fit absorbed thermal plasma models to the spectra from each annulus. These models are able to produce temperature $kT$ and intrinsic column density $N_{\rm H}^{1266}$ which we show in the first two panels in the top left of Figure~\ref{fig:gradients}. Positive values of distance are of the northern outflow, and negative are along the southern outflow. We find the column densities peak in the northern outflow with \edit1{$N^{\rm{1266}}_{\rm{H}}=(1.2^{+0.26}_{-0.20})\times10^{22}\;\rm{cm^{-2}}$ at $\sim$386~pc} from the center and are at their lowest, \edit1{$N^{\rm{1266}}_{\rm{H}}=(4.7^{+2.1}_{-1.3})\times10^{21}\;\rm{cm^{-2}}$}, at $\sim$$-$446~pc. We find the hot gas temperature peaks at \edit1{$kT = 1.8^{+2.1}_{-0.54}$~keV or $(2.09^{+2.44}_{-0.63})\times10^7\;\rm{K}$} in the first northern region 205~pc from the center, and it decreases with distance from the AGN, particularly in the southern outflow. \edit1{The outflow temperatures are consistent with those reported in \cite{Su2015} and adopted by \cite{Eskenasy2024}, where $kT=0.86\pm0.09$~keV.}

From the thermal plasma model (APEC), the norm parameter can provide an estimate of the electron number density ($n_{\rm e}$) as it relates to the emission measure, $\mathrm{norm}=(10^{-14}\rm{EM})/(4\pi D^2)$, where $\mathrm{EM=\int n_{\rm e} n_{\rm H} dV}$. Assuming $n_{\rm e}=1.2 n_{\rm H}$, and a spherical shell geometry for each region in Figure~\ref{fig:xray_image}, we plot the $n_{\rm e}$ in the third top panel of Figure~\ref{fig:gradients}. We also assume a filling factor $f$ of unity. We find the density peaks at 205~pc with a value of high value of $4.2\;\rm{cm^{-3}}$ and decreases with distance from the AGN. We compare this value to that of other local X-ray emitting outflows in Section~\ref{sec:comparisons}. 

Adopting the derived values of $n_{\rm e}$, we also calculate the cooling time of the hot gas $t_{\rm cool}$, its thermal pressure $P/k$, and thermal energy $E_{\rm th}$. The cooling time is calculated as $t_{\rm cool}=3kT/\Lambda n_{\rm e}$, where $\Lambda$ is the cooling function (with units of erg cm$^{3}$ s$^{-1}$) for a solar plasma using \textsc{PyAtomDB} \citep{PyATOMDB}. We find the cooling times to be short at around 1~Myr. The thermal pressures are calculated as $P/k=2n_{\rm e} T$, peaking $\sim10^8\;\mathrm{K\;cm^{-3}}$ and decreasing with distance to $\sim10^6\;\mathrm{K\;cm^{-3}}$. As performed for the Taffy Galaxies \citep{Appleton2015}, we derived the thermal energy as $E_{\rm th} = 3/2n_{\rm e} VkT$, where $V$ is the spherical shell volume. We find the thermal energy to be between $E_{\rm th} \approx 9\times10^{54}-1\times10^{56}$~erg. \edit1{These quantities are all shown in Figure~\ref{fig:gradients} and listed in Table~\ref{tab:ngc1266_gradients}.}

\begin{figure}
    \centering
    \includegraphics[width=0.47\textwidth]{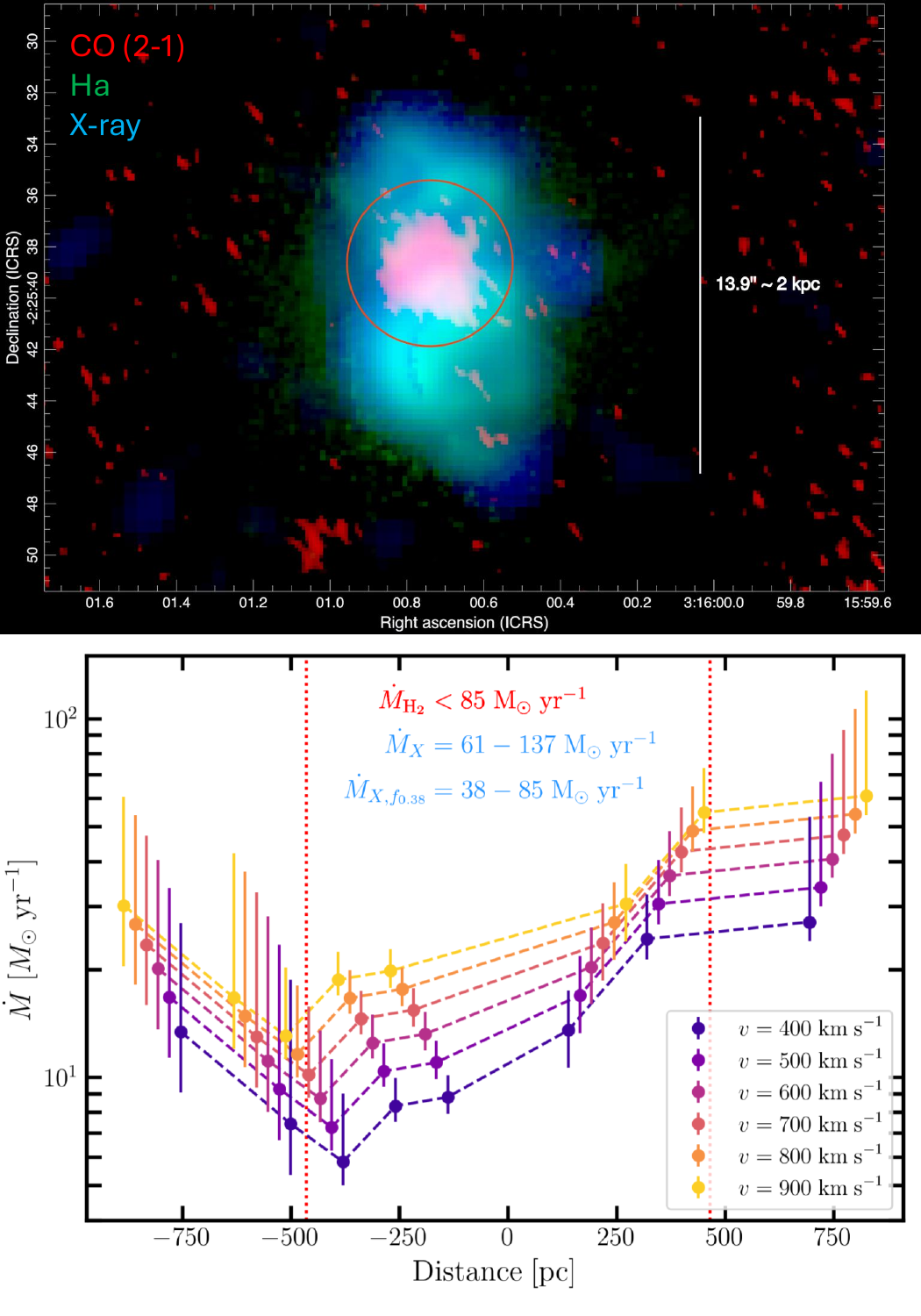}
    \caption{\edit1{\textit{Top:} Three-color image of NGC~1266 where red is the SMA CO(2-1) observations from \cite{Alatalo11}, green is VLT/MUSE H$\alpha$, and blue is \textit{Chandra} broadband X-ray emission. The red circle indicates the extent of the CO emission found in \cite{Alatalo11}, 3.23\arcsec{} or $\approx460$~pc. \textit{Bottom:} Estimated hot gas mass outflow rates using the mass values in Table~\ref{tab:ngc1266_gradients} and different velocities shown by the different colored lines ranging from $400-900\;\mathrm{km\;s^{-1}}$, the velocities of the warm gas found by \cite{Davis2012}. The vertical dotted red lines show the extent of the CO(2-1) emission along with the estimated molecular gas outflow rates, the X-ray estimated outflow rates within the same region, and the X-ray outflow rates if the filling factor were low enough to keep the hot gas mass outflow rate under the molecular gas estimate as is commonly expected \citep{Veilleux2020}. }}
    \label{fig:massoutflow}
\end{figure}

In the bottom right panel of Figure~\ref{fig:gradients}, we show the mass gradient of the X-ray outflow. We calculate the outflow mass as $M_{\rm{out}}=n_{\rm H} Vm_{\rm H}$, giving values ranging from $M_{\rm out} \approx 4\times10^6-5\times10^7\;\rm{M_\odot}$ totaling \edit1{$8.0\times10^{7}\;\rm{M_\odot}$} in the north and \edit1{$5.4\times10^{7}\;\rm{M_\odot}$} in the south. \edit1{To derive mass outflow rates, we use a range of outflow velocities from \cite{Davis2012}, who observed ionized gas velocities near zero around the AGN up to a maximum of 900~km~s$^{-1}$. Direct measurements of the hot gas kinematics are non-existent (though with XRISM more exceptions are being made), thus we use the optical results as a lower estimate since we expect the hot gas to exceed the warm phase in velocity \citep{Thompson2024}. We evaluate the mass outflow rate as $\dot{M}_X = M_{\rm out}\,v_{\rm out}/r$, where $r$ is the projected distance of each annulus from the AGN, and sum over all annuli within the extent of the molecular outflow ($r \lesssim 460$~pc; \citealt{Alatalo11}). We show these hot gas mass outflow rates in Figure~\ref{fig:massoutflow}, where we plot different gradients based on varying velocities. We find that within this region, the hot gas mass outflow rates are comparable to or exceed the molecular outflow rate: for the fastest velocity of $v_{\rm out}=900\;\rm{km\;s^{-1}}$, the hot gas outflow rate exceeds the molecular gas outflow rate by a factor of 1.6, where the molecular gas estimates are from \cite{Otter2026}. The minimum and maximum outflow rates of $\dot{M}_X=61$ and $137\;\rm{M_\odot\;yr^{-1}}$ correspond to velocities of $v_{\rm out}=400$ and $900\;\rm{km\;s^{-1}}$, respectively. We note that the filling factor is uncertain, so these values carry a factor of $f^{1/2}$. A filling factor of $f=0.38$ would bring the hot gas outflow mass into agreement with the molecular outflow mass at the fastest velocity. The spatial coincidence of the X-ray emission with the ionized gas outflow \citep{Davis2012} provides the primary evidence for outward motion of the hot gas. We discuss possible interpretations of this finding, as well as our assumptions on filling factor and warm-hot phase coupling, in Section~\ref{ssec:prev_work}.}

%To derive outflow rates, we use the outflow velocities from \cite{Davis2012} who observed southern  velocities from near zero around the AGN to a maximum of 900~km~s$^{-1}$ and northern velocities of 600~km~s$^{-1}$. If we assume the wind steadily and equally increases velocity in each shell  shown in Figure~\ref{fig:xray_image} and that the warm and hot gas phases are coupled, we find the mass outflow rate in the north is 62.7~$\rm{M_\odot/yr}$ and in the south is 54~$\rm{M_\odot/yr}$, leading to a combined total mass outflow rate of 116.7~$\rm{M_\odot/yr}$. Including only the regions exceeding NGC~1266's escape velocity of 340~km~s$^{-1}$ \citep{Alatalo11}, we find a total escape mass outflow rate of about 99~$\rm{M_\odot/yr}$. This estimate exceeds that found in \cite{Alatalo11} with CO gas (13~$\rm{M_\odot/yr}$) and in Otter et al. (2025 submitted) with HCN (85~$\rm{M_\odot/yr}$). We discuss possible interpretations of this finding as well as consider our assumptions on filling factor and warm-hot phase coupling in Section~\ref{ssec:prev_work}.

\begin{deluxetable*}{lcccccccccr}
\tablecolumns{11}
\tablewidth{0pt}
\tablecaption{NGC~1266 Annuli Best-fit Properties\label{tab:ngc1266_gradients}}
\tablehead{
\colhead{Region} &
\colhead{Distance} &
\colhead{$N_{\mathrm{H}}$} &
\colhead{$kT$} &
\colhead{$n_e$} &
\colhead{$t_{\mathrm{cool}}$} &
\colhead{$P/k$} &
\colhead{$E_{\mathrm{th}}$} &
\colhead{$L_X$} &
\colhead{$M_{\mathrm{out}}$} & 
\colhead{$\chi^2$/d.o.f} \\
\colhead{} &
\colhead{(pc)} &
\colhead{$(10^{22}\,\mathrm{cm^{-2}})$} &
\colhead{(keV)} &
\colhead{$(\mathrm{cm^{-3}})$} &
\colhead{(Myr)} &
\colhead{$(10^{7}\,\mathrm{K\,cm^{-3}})$} &
\colhead{$(10^{55}\,\mathrm{erg})$} &
\colhead{$(10^{39}\,\mathrm{erg\,s^{-1}})$} &
\colhead{$(10^{6}\,M_\odot)$} &
\colhead{}
}
\startdata
S5 & $-$820 & $0.58_{-0.10}^{+0.45}$ & $0.25_{-0.06}^{+0.07}$ & $0.33_{-0.11}^{+0.33}$ & $1.2_{-0.49}^{+1.3}$ & $0.19_{-0.08}^{+0.20}$ & $1.9_{-0.77}^{+2.0}$ & $9.6_{-2.1}^{+2.1}$ & $27_{-8.7}^{+27}$ & {17/12}\\
S4 & $-$567 & $0.66_{-0.20}^{+0.35}$ & $0.36_{-0.11}^{+0.14}$ & $0.86_{-0.24}^{+1.3}$ & $0.80_{-0.33}^{+1.3}$ & $0.71_{-0.29}^{+1.1}$ & $1.1_{-0.43}^{+1.7}$ & $9.9_{-0.57}^{+0.57}$ & $10_{-2.9}^{+16}$ & {19/24} \\
S3 & $-$446 & $0.47_{-0.13}^{+0.21}$ & $0.60_{-0.19}^{+0.10}$ & $0.85_{-0.12}^{+0.46}$ & $1.3_{-0.43}^{+0.72}$ & $1.2_{-0.41}^{+0.68}$ & $1.1_{-0.37}^{+0.63}$ & $4.9_{-0.40}^{+0.40}$ & $6.4_{-0.89}^{+3.5}$ & {25/25}  \\
S2 & $-$326 & $0.99_{-0.10}^{+0.18}$ & $0.66_{-0.07}^{+0.11}$ & $1.6_{-0.16}^{+0.33}$ & $0.70_{-0.10}^{+0.18}$ & $2.5_{-0.36}^{+0.66}$ & $1.3_{-0.18}^{+0.33}$ & $13_{-0.99}^{+0.99}$ & $6.6_{-0.63}^{+1.3}$ & {43/38}  \\
S1 & $-$205 & $1.0_{-0.12}^{+0.16}$ & $0.70_{-0.08}^{+0.08}$ & $2.7_{-0.28}^{+0.41}$ & $0.45_{-0.07}^{+0.09}$ & $4.5_{-0.70}^{+0.85}$ & $0.89_{-0.14}^{+0.17}$ & $11_{-0.65}^{+0.65}$ & $4.4_{-0.45}^{+0.67}$ & {20/31}  \\
N1 & 205 & $1.0_{-0.81}^{+0.86}$ & $1.8_{-0.54}^{+2.1}$ & $4.2_{-0.90}^{+1.2}$ & $1.6_{-0.57}^{+1.9}$ & $18_{-6.5}^{+22}$ & $3.5_{-1.3}^{+4.3}$ & $2.2_{-0.30}^{+0.30}$ & $6.8_{-1.5}^{+2.0}$ & {7.4/10}  \\
N2 & 386 & $1.2_{-0.20}^{+0.26}$ & $0.85_{-0.22}^{+0.13}$ & $2.0_{-0.25}^{+0.66}$ & $0.78_{-0.22}^{+0.28}$ & $4.0_{-1.1}^{+1.4}$ & $5.6_{-1.6}^{+2.0}$ & $4.3_{-1.3}^{+1.3}$ & $23_{-2.9}^{+7.6}$ & {26/20}  \\
N4 & 760 & $0.58_{-0.13}^{+0.79}$ & $0.76_{-0.22}^{+0.19}$ & $0.54_{-0.06}^{+0.52}$ & $2.5_{-0.80}^{+2.5}$ & $0.94_{-0.30}^{+0.94}$ & $11_{-3.5}^{+11}$ & $3.4_{-0.48}^{+0.48}$ & $50_{-5.8}^{+49}$ & {26/26}  \\
\enddata

\end{deluxetable*}

%\begin{figure}
%    \centering
%    \includegraphics[width=0.45\textwidth]{NGC1266_South_2apec_acx.pdf}
%    \includegraphics[width=0.45\textwidth]{NGC1266_South_2apec_noacx.pdf}
%    \caption{Spectra extracted from the composite southern region from observation 19498. \textit{Top}: The CX emission is in blue (ACX), the warm-hot thermal plasma component (APEC) is in green, and the hot thermal plasma component is in blue. The data points are in black crosses, and the residuals to the best-fit model are given below. We find that in the south, a significant portion of the broad-band X-ray emission ($\sim38$\%) is dominated by CX emission. \textit{Bottom}: Same as the top panel but the model does not include CX. Without CX, the model has larger residuals at soft X-ray energies and yields a statistically worse fit.}
%    \label{fig:cx}
%\end{figure}

Using F-tests, we found that CX was statistically necessary in the southern regions but not in the north of NGC~1266. In Figure~\ref{fig:cx}, we plot the spectra from observation 19498 as an example to show the model with and without a CX contribution. We find that the CX emission contributes 38\% of the total luminosity, while the CIE plasma (\textsc{APEC}) components for the warm-hot and hot phase contribute 50\% and 12\%, respectively. Thus over a third of the X-ray emission of the southern lobe is a result of interactions between the hot wind and cooler surrounding medium. We note that the non-detection of CX in the north does not mean these interactions are not taking place but rather may be a result of the north pointing away from the observer as shown by the CO and ionized gas kinematics in \cite{Alatalo11}  and \cite{Davis2012}, respectively. This approaching versus receding outflow asymmetry is also observed in NGC~253 \citep{Lopez2023} and NGC~4945 \citep{Porraz2024} where the receding outflows do not have CX detections, likely because of extinction attenuating the soft X-ray emission. We also find this in Figure~\ref{fig:cx} where the north has lower emission in the soft X-ray ($<2$~keV) range compared to the south.

\subsection{The Warm Outflow Phase}

\begin{figure*}
    \centering
    \includegraphics[width=0.92\textwidth]{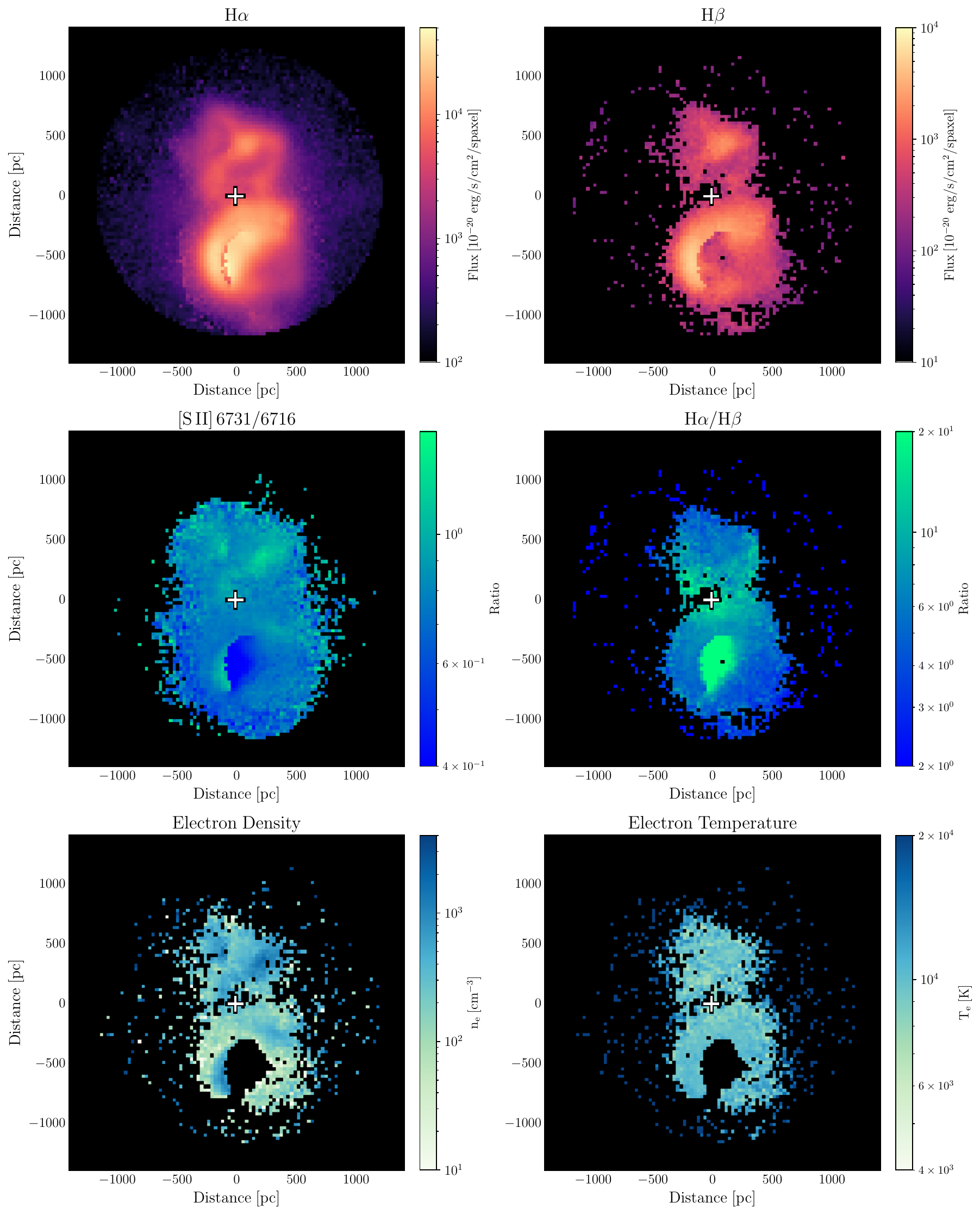}
    \caption{MUSE data products used in this paper, all of which were clipped to a signal-to-noise ratio of 5. \textit{Top Row}: MUSE H$\alpha$ and H$\beta$ flux maps of NGC~1266. \textit{Middle Row}: Sulfur doublet ratio and H$\alpha$/H$\beta$ ratio, where the former shows it hits the low density limit in the cavity, and the latter implies heavy extinction. \textit{Bottom Row}: Electron density $n_{\rm e}$ map from the sulfur doublet and electron temperature map $T_{\rm e}$ from the \ion{N}{2} lines. Both values are undefined in the region of the cavity due to the low density limit. }
    \label{fig:MUSE}   
\end{figure*}

In Figure~\ref{fig:MUSE}, we present several optical line images, ratios, and derived quantities. All images were masked to have a signal-to-noise ratio of $\mathrm{SNR}>5$ using the flux error maps produced by the PHANGS-MUSE pipeline. As shown in Figure~\ref{fig:3color}, the H$\alpha$-emitting gas follows a similar morphology to the X-ray emission, with brightness enhancements in the left, south side of the outflow and in the northern, right side. Due to lower signal in the H$\beta$ map, it is less extended but nonetheless has similar features as in the H$\alpha$ image. Interestingly and more evident than in the H$\alpha$ image is a cavity-like structure to the south of the AGN location (with the latter marked by a white cross). The H$\alpha$/H$\beta$ ratio highlights the extremes of the two wavelengths in the middle left panel of Figure~\ref{fig:MUSE}. The high ratio is indicative of extreme extinction and when calculating $A_{\rm V}$ using the \cite{Cardelli1989} formulation, we find values of $A_{\rm V}\sim10$, though we note that this assumes a screen-like geometry for the dust along the line of sight. 

In Figure~\ref{fig:MUSE}, we also show the sulfur doublet ratio of the $6731$~\AA{} and $6716$~\AA{} in the middle left panel that serves as a density diagnostic. 
Once again, we find that the cavity is evident in the line ratio and even reaches the low density limit. 
This is further confirmed in the bottom left panel where the warm gas electron density $n_{\rm e}$ is undefined in the cavity. 
We derived electron densities and temperatures for each pixel using PyNeb \citep{pyneb}, which simultaneously solves for $n_{\rm e}$ and $T_{\rm e}$ from the \ion{S}{2} doublet and \ion{N}{2} $5755$ \AA{} / $6548+6584$ \AA{} ratio.
We find $n_{\rm e}$ values between $10-3000\;\rm{cm^{-3}}$ and $T_{\rm e}$ values of $(0.4-2)\times10^4$~K. The resulting thermal pressures $P/k$ are thus $4\times10^{3}-6\times10^7\;\rm{K\;cm^{-3}}$. 

\begin{figure}
    \centering
    \includegraphics[width=0.40\textwidth]{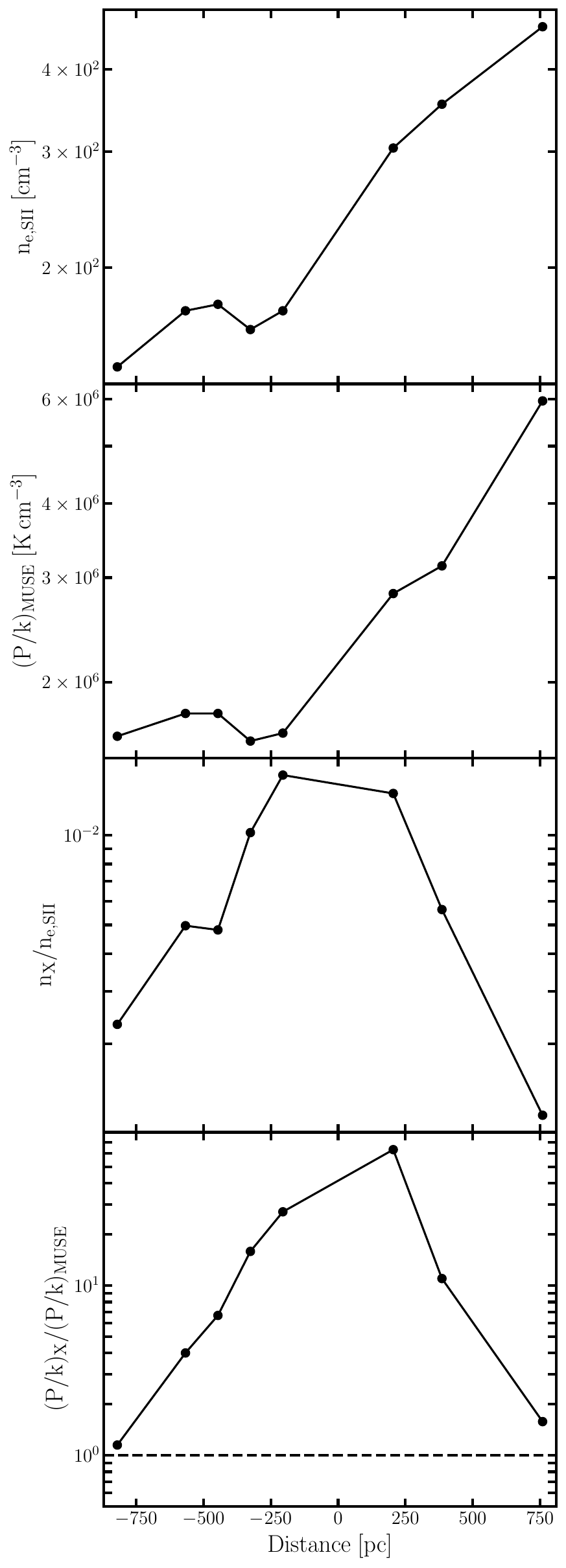}
    \caption{\edit1{\textit{First and Second:} Gradients with distance of the median $n_{\rm e}$ and $P/k$ values from the MUSE data. \textit{Third and Fourth:} Ratio plots comparing the $n_{\rm e}$ and $P/k$ values from MUSE and our X-ray spectral analysis. We find that the warm and hot outflow phases reach near pressure equilibrium at the outflows maximum extent.}}
    \label{fig:ratios}
\end{figure}

In Figure~\ref{fig:ratios}, we plot the median $n_{\rm e}$ and $P/k$ values from the MUSE data and their ratios with the X-ray constraints. For the MUSE data, we overlaid the same regions from the X-ray spectral extractions (as defined in Figure~\ref{fig:xray_image}), and we took the median values from the $n_{\rm e}$ and $P/k$ maps in each region. The $n_{\rm e}$ and $P/k$ profiles are higher in the north than in the south though overall similar orders of magnitude, unlike the X-ray that drops more steeply with distance (Figure~\ref{fig:gradients}). In the ratio plots, as expected, we find the warm gas $n_{\rm e}$ values to be greater than the X-ray's by two to three orders of magnitude highlighting the more diffuse nature of the hot gas in NGC~1266. When comparing the pressures, we find that the X-ray emitting gas is overpressurized compared to the warm phase by about an order of magnitude. However, as the wind approaches the maximum observed extent of the X-ray emission ($>700$~pc), the two phases reach near pressure equilibrium.  

\section{Discussion} \label{sec:disc}

\subsection{Comparison to Previous Work} \label{ssec:prev_work}

Since the discovery of NGC~1266's unusually large molecular gas reservoir in the ATLAS$^{\rm{3D}}$ \citep{Cappellari2011} survey, numerous works have analyzed the outflow of the system across the electromagnetic spectrum. \cite{Alatalo11} first analyzed in depth the CO data from the CARMA and SMA and discovered the molecular gas has both a highly concentrated nuclear structure and a diffuse, envelope-like extended component. The former is likely the disk that constricts the outflow, and the latter is the outflowing molecular gas itself, which has an average velocity of about 177~km~s$^{-1}$. The molecular outflows extend about $R_{\rm{out}}\sim460$~pc, which is smaller than the H$\alpha$ and X-ray that both go out to roughly $R_{\rm{out}}\sim1$~kpc \edit1{(see Figure~\ref{fig:massoutflow})}. According to \cite{Davis2012}, this means that either the mm observations were not sensitive enough or the CO is being destroyed at these larger distances. They also noted the location of the AGN coincides with the hard X-ray emission in the \textit{Chandra} data, which we confirm with deeper observations shown in the Figure~\ref{fig:xray_image}. 

\cite{Davis2012} studied the warm gas phase of NGC~1266's outflow using SAURON and GMOS IFU data. They found a much more rapid outflow component than the molecular gas, reaching velocities of $v_{\rm{out}}\approx900$~km~s$^{-1}$. These velocities are similar to the shock velocities derived by fitting the models of \cite{Allen2008} to various line ratios, indicating that the ionized gas emission is predominately or fully from shock ionization. The morphology of the optical line maps in \cite{Davis2012} match the X-ray images closely, as shown in Figure~\ref{fig:3color}. This can be interpreted as the hot gas shocking the cooler ISM and producing the optical emission. As shown in Figure~\ref{fig:cx}, at least in the southern lobe, over a third of the X-ray emission is a result of CX, providing additional evidence for significant hot-cool phase interactions in NGC~1266's outflow. 

%Absent from the \cite{Davis2012} GMOS images is the cavity in the southern outflow we observed in Figure~\ref{fig:MUSE}. 

The GMOS images presented in \cite{Davis2012} are missing the cavity in the southern outflow apparent in the MUSE data in Figure~\ref{fig:MUSE}. Although \cite{Davis2012} found the \ion{S}{2} doublet ratio reaches the low density limit in several parts of the southern outflow, they did not resolve the same structure. Also studying IFU data, \cite{Eskenasy2024} analyzed MUSE images from several early-type galaxies including the same data we use for NGC~1266, and they detected the cavity-like structure. They presented the H$\alpha$ velocity and velocity dispersion maps, showing that the cavity is redshifted in contrast to the rest of the southern outflow that is blueshifted. If we consider the redshifted velocity structure of the cavity along with it hitting the low density limit of \ion{S}{2} doublet diagnostic, we surmise that the cavity is a real, astrophysical feature and potentially revealing the opposite side of the outflow cone. If this is the case, then the longer path length through the outflow, along with the dusty nature of NGC~1266's nucleus (see the HST images in \citealt{Nyland2013}), would explain the high levels of extinction highlighted by the large H$\alpha$/H$\beta$ ratios in Figure~\ref{fig:MUSE}.

Previous work on NGC~1266 derived the HI column densities of the outflow using 1.4~GHz continuum observations in \cite{Alatalo11}, \cite{Davis2012}, and \cite{Nyland2013}, and they ranged from $N_{\rm H}^{1266} \approx (1.2-2.4)\times10^{21}\;\rm{cm^{-2}}$. \cite{Alatalo2015} fit the archival \textit{Chandra} and XMM-Newton available at the time and derive $N_{\rm H}^{1266}=(1.5\pm0.2)\times10^{21}\;\rm{cm^{-2}}$, in agreement with the 1.4 GHz continuum derived HI column densities. Our deeper \textit{Chandra} observations find $N_{\rm H}^{1266}$ to range from \edit1{$(4.7^{+2.1}_{-1.3})\times10^{21}$ to $(1.2^{+0.26}_{-0.20})\times10^{22}\;\rm{cm^{-2}}$}, similar order of magnitude albeit slightly higher. 

Lastly, and perhaps most surprisingly, we find that the hot phase outflow rates exceed the molecular gas constraints from \edit1{\cite{Otter2026} assuming a maximum observed warm gas velocity of $900\;\rm{km\;s^{-1}}$ from \cite{Davis2012}}. In typical galactic outflows, the cold phases dominate the mass budget \citep{Veilleux2005,Veilleux2020}. For the hot phase to exceed the cold phase in outflow rate, a few explanations should be considered. First, the outflow may be extremely mass-loaded, resulting in higher-than-typical densities. Most hot winds have $n_{\rm e} \sim 0.01-1\;\rm{cm^{-3}}$ (see Table~\ref{table:comparison}), whereas we observe peak values closer to about $4\;\rm{cm^{-3}}$. The substantial charge exchange (CX) emission in the outflow may support this as it suggests the hot wind is interacting with abundant cool material and could entrain it.

Second, the mass outflow rates derived in this paper are likely overestimates. Measurements of outflow rates in edge-on systems are notoriously difficult, especially without resolved kinematics (like in the hot phase), as projection effects can bias the inferred geometry, filling factor, and luminosity, particularly due to limb-brightening along cone edges. If the filling factor is smaller than unity or the volume decreased, this would raise the implied densities and decrease the mass since $M_{\rm{out}}=n_{\rm H} m_{\rm H} f^{1/2}V$. Thus, our estimated $n_{\rm e}$ and $M_{\rm{out}}$ are lower and upper limits, respectively. \edit1{If we set the HCN values (85~$\rm{\dot M_{\rm out}}$) as upper limits, then we find the filling factors for the hot gas to be 0.38.} Such filling factors would make the hot phase outflow rates consistent with the cool phase (albeit still unusual); however given the uncertainties, we cannot conclusively determine which considerations are more probable, and thus we note that our hot phase mass outflow rates are order-of-magnitude estimates.

%However, we emphasize that the volumes assumed (spherical shells) and the adopted filling factor (unity) already favor lower density, and therefore lower mass, estimates. If the true filling factor is smaller than unity, this would raise the implied densities and increase the mass, exacerbating the discrepancy with the CO-derived values. Given these uncertainties, we cannot conclusively determine which theory is most probable, and thus consider our hot phase mass outflow rates to be an order of magnitude estimate.

\subsection{What Drives the Outflow?}

As mentioned earlier, Figure~\ref{fig:ratios} demonstrates that the X-ray and optical emitting phases reach pressure equilibrium at the maximum extent of the outflow, $\sim$750~pc. In Figure~\ref{fig:gradients}, we also find the cooling times are of order 1~Myr. We can estimate an advection time for the outflow as $t_{\rm{adv}}=R/v_{\rm{out}}$ using the velocity of \cite{Davis2012} of $v_{\rm out} = 900$~km~s$^{-1}$ and the maximum outflow extent of $R \approx1$~kpc. We find that the advection time is $t_{\rm adv} \approx 1.1$~Myr, comparable to the cooling times, suggesting that the outflow is likely undergoing bulk radiative cooling \citep{Wang1995,Thompson2016}. This finding implies that the hot wind, barring future AGN episodes, is no longer able to maintain a strong pressure gradient to drive further expansion. Instead, it may be entering a mechanically-exhausted phase in which the outflow stalls and the hot gas begins to cool and mix with the surrounding medium, potentially explaining the close H$\alpha$ correspondence \citep{Eskenasy2024}. We do exercise caution here with this interpretation as the uncertainties with the volume and filling factor may result in the densities being underestimated. If that is the case, then the X-ray derived pressures would increase above pressure equilibrium, thereby continuing the outflow episode.

Whether or not the outflow has stalled, the origin of the energy driving it remains uncertain. \edit1{The physical picture is that the radio jet, which \cite{Nyland2013} show is deeply embedded within the dense nuclear molecular gas reservoir, inflates a cocoon of hot plasma that drives a forward shock into the surrounding ISM \citep{Wagner2012,Mukherjee2016}. This overpressured hot cocoon acts as the primary intermediary between the jet and the cold molecular gas, driving the bulk of the molecular outflow as it expands. In this frustrated jet scenario \citep{Mukherjee2016}, where the jet remains confined on sub-kpc scales, the coupling is particularly efficient as essentially all of the jet energy is deposited locally rather than escaping to larger scales.} \cite{Nyland2013} found that the minimum jet energy is about $1.7\times10^{54}$ erg, which is an order of magnitude lower than the kinetic molecular energy estimated by \cite{Alatalo11} at $10^{55}$ erg. The \edit1{former} argues that since the true energy of the radio lobes may be higher than the minimum estimate, it is feasible that the lobes are capable of driving the outflow. In Figure~\ref{fig:gradients}, we show the thermal energy distribution across the outflow, finding values that range from $\sim10^{55}-10^{56}$~erg. Our results indicate that the hot, X-ray-emitting phase contains a substantial fraction of the total energy budget of the outflow. If the true jet energy exceeds the minimum estimate from \cite{Nyland2013}, then the observed thermal energy in the hot phase implies that a non-negligible fraction of the jet's power must have been deposited into the surrounding ISM. \edit1{While the exact coupling efficiency remains uncertain due to the unknown total jet energy, the hot gas carries a substantial portion of the outflow energy budget in both thermal and kinetic form. The thermal energy of the hot phase ranges from $\sim10^{55}$--$10^{56}$~erg, while the kinetic energy $E_{\rm kin} = \frac{1}{2}M_X v_{\rm out}^2 \sim 10^{56}$--$10^{57}\,f^{1/2}$~erg for $v_{\rm out}=400$--$900$~km~s$^{-1}$, comparable to or exceeding the thermal energy at higher velocities and comparable to the molecular outflow kinetic energy of $10^{55}$~erg \citep{Alatalo11}.}

%If the radio lobes are responsible for driving the outflow, another reason the jet energy derived by \cite{Nyland2013} must be higher relates to the pressure balance. Using their reported lobe radii of $r_{\rm{south}} = 0.34$~kpc and $r_{\rm{north}} = 0.12$~kpc, we compute the radio lobe pressures via the plasma pressure equation, $p_{\rm{min}} = u_{\rm{min}}/3 = U_{\rm{min}} / 3V$, where $U_{\rm{min}}$ is the minimum jet energy and $V$ is the lobe volume assuming spherical geometry. We find $p_{\rm{north}} = 2.66 \times 10^{-9}\;\rm{erg~cm^{-3}}$ and $p_{\rm{south}} = 1.17 \times 10^{-10}\;\rm{erg~cm^{-3}}$. For comparison, the X-ray pressure ranges from $1.3 \times 10^{-9}$ to $2.5\times10^{-8}\;\rm{erg~cm^{-3}}$ in the north and from $2.6 \times 10^{-10}$ to $1.7 \times 10^{-9}\;\rm{erg~cm^{-3}}$ in the south. These results show that the radio lobes are underpressurized relative to the hot gas near the nucleus, and they only reach pressure equilibrium at the maximum extent of the outflow. Thus to exceed both the X-ray pressure and the energy constraints, the radio jet must have $U\gtrsim10^{56}$~erg, two orders of magnitude higher than the \cite{Nyland2013} minimum energy constraint.

\subsection{Comparison to other wind hosts}
\label{sec:comparisons}

\begin{deluxetable*}{lccccc}
\tablecolumns{6}
\tablewidth{0pt}
\tablecaption{Outflow Properties of Comparison Galaxies \label{table:comparison}}
\tablehead{
\colhead{Galaxy} & 
\colhead{Type} & 
\colhead{$kT$ [keV]} & 
\colhead{$\rm{n_e}$[cm$^{-3}$]} & 
\colhead{$P/k$ [$\times10^6$ K cm$^{-3}$]} & 
\colhead{Reference}
}
\startdata
NGC~1266\tablenotemark{a}       & AGN & $0.23-0.77\;[0.24-1.8]$     & $0.24-0.73\;[0.33-4.2]$     & $1.9-11\;[1.9-180]$  & This Paper \\
M84            & AGN & $0.5-0.9$     & $0.1-0.5$     & $6-20$   & \cite{Bambic2023} \\
NGC~3801       & AGN & $0.2-1.0$ &  $0.01-0.03$ & $0.04-1.2$ & \cite{Croston2007} \\
NGC~6764       & AGN & $0.6-0.8$     & $0.2-0.3$     & $4.2-8.4$ & \cite{Croston2008} \\
Circinus       & AGN & $0.7-1.8$    & $0.02$ & $0.8-1.4$ & \cite{Mingo2012} \\
Mrk~573        & AGN & $0.76-1.29$  & $0.14-0.52$   & $2.75-16.9$ & \cite{Paggi2012} \\
M82            & SB  & $0.4-3.4$     & $0.04-1.8$    & $0.6-212$ & \cite{Lopez2020}, \\
               &     &               &               &           & \cite{Nguyen2021} \\
NGC~253        & SB  & $0.2-0.9$     & $0.1-2$       & $0.8-62$  & \cite{Lopez2023} \\
NGC~4945       & SB  & $0.1-1$       & $0.04-1.7$    & $0.2-42$  & \cite{Porraz2024} \\
\enddata
\tablenotetext{a}{Values listed for NGC~1266 are for the southern and northern lobe to best compare to previous AGN studies that use similar geometry. Since these values have two temperature components, we take the flux weighted average for each component and report it in the table. The values in brackets are the values from Figure~\ref{fig:gradients}. Compared to the rest of the sample, we find that NGC~1266 either sits at the high end of ranges or is a factor of a few denser than other X-ray emitting winds.}

\end{deluxetable*}

Compared to other X-ray emitting wind hosts, NGC~1266 sits in a unique position, particularly due to its high density. We note that comparing $n_{\rm e}$ across different systems can introduce uncertainties due to differing volumes. To best compare to other AGN radio jet driven systems, we rederive the properties for NGC~1266 for the total southern and northern outflows using the spectral models in Table~\ref{table:specfitscx}. For comparison, we also include starburst-driven winds \citep{Lopez2020,Lopez2023,Porraz2024}. \edit1{We compile the values in Table~\ref{table:comparison}, including the re-derived integrated properties of the northern and southern outflow lobes of NGC~1266. The comparison sample was selected to include all available galaxies with spatially resolved X-ray spectroscopic measurements of outflowing gas in the literature, spanning both AGN jet-driven and starburst-driven winds, reflecting the currently limited number of such studies.}

First, we compare our total outflow shell results for NGC~1266 to previous \textit{Chandra} X-ray derived properties of AGN hosts: M84 \citep{Bambic2023}, NGC~3801 \citep{Croston2007}, NGC~6764 \citep{Croston2008}, the Circinus Galaxy \citep{Mingo2012}, \edit1{and Mrk 573 \citep{Paggi2012}}. We find that that NGC~1266, even after calculating $n_{\rm e}$ for a larger volume, still has $n_{\rm e}$ values two orders of magnitude greater than NGC~3801 and Circinus. Meanwhile with M84, \edit1{Mrk 573,} and NGC~6764, the $n_{\rm e}$ values are similar, though NGC~1266 is still on the higher end of the ranges. The temperatures are all roughly consistent, as expected of the hot wind phase. For NGC~3801, NGC~6764, and Circinus, the total thermal energies span the range of $0.5-5\times10^{55}$~erg, also roughly consistent with NGC~1266. Thus in this small sample, NGC~1266 is similarly energetic to other AGN though above average in hot wind densities. More follow-up work is needed with larger samples to truly understand where NGC~1266 lies compared to other AGN outflows.

While NGC~1266 is not a starburst \citep{Nyland2013} as its star formation rate is too low to drive a wind \citep{Otter2024}, we still compare the outflow to other previous edge-on, spatially-resolved X-ray analysis of starburst wind hosts: M82 \citep{Lopez2020}, NGC~253 \citep{Lopez2023}, and NGC~4945 \citep{Porraz2024}. For these sources, the spectra were modeled for smaller regions as opposed to the total lobes as is done for the AGN studies. As a result, the values shown in brackets in Table~\ref{table:comparison} are most comparable between NGC~1266 and the starbursts. We find the values for NGC~1266 to be well within the range of the starburst galaxies except for $n_{\rm e}$ in the inner $\pm325$ that exceeds the starbursts. This highlights the incredibly dense nature of NGC~1266's wind likely due to its massive molecular gas reservoir being entrained by the hot phase and creating the CX emission. 

%More follow-up work is needed in comparing AGN and starburst driven outflows to find their differences (and similarities) and what causes them.  

%Therefore, our comparisons indicate that while NGC~1266 is as hot and energetic as other local AGN jet hosts, it is incredibly more dense than the others, implying heavy mass loading that is consistent with the detection of CX and the evidence of radiative cooling.

%It may be that the starburst systems have higher mass loading than the AGN as a result of the wind being driven by supernovae feedback in dense clusters. This may be supported by the fact that CX is detected in each of the starburst outflows, though we note the AGN studies did not look for such emission. Considering the wealth of molecular gas in NGC~1266's nucleus, it may be it has entrained more material than the other AGN systems listed in Table~\ref{table:comparison}. More follow-up work is needed in comparing AGN and starburst driven outflows to find their differences (and similarities) and what causes them.  

\subsection{Charge Exchange Detection}

Previous work has shown that charge exchange (CX), the stripping of an electron from a neutral atom by an ion, can be a prominent feature in X-ray emission, though much of the work has been for star-forming systems rather than AGN. For example, \cite{Liu2012} showed that the K$\alpha$ triplet of He-like ions required a CX component to account for the observed line ratios for several nearby star-forming galaxies. The recent works of \cite{Lopez2020}, \cite{Lopez2023}, and \cite{Porraz2024} also found significant CX contributions ($\sim$12-42\%) to broad-band \textit{Chandra} spectra in M82, NGC~253, and NGC~4945. 

CX analyses for AGN-driven outflows are scarce, though one example is the work of \cite{Yang2020} who analyzed XMM-Newton RGS spectra of M51's central region. They found that most of the soft X-ray emission originates from the AGN outflow and its interactions with the neutral material surrounding it. Due to the observations using X-ray gratings, they had high spectral resolution to observe directly the prominent \ion{O}{7} forbidden lines that are enhanced with CX. They found that CX makes up a significant fraction of the diffuse X-ray emission, about 21\%. The robust result of \cite{Yang2020} indicates that AGN-driven outflows, like the starburst ones, are able to produce CX emission as long as neutral material is interacting with hot gas, as is the case in NGC~1266. Unfortunately, high resolution X-ray spectra of NGC~1266, particularly of the \ion{O}{7} forbidden lines, are not available, so future observations are merited to better quantify the CX contribution. Moreover, the detection of CX in both M51 and NGC~1266 provides a case for revisiting the AGN outflow studies listed in Table~\ref{table:comparison} to test for CX emission.

\section{Conclusion} \label{sec:conclusion}

We analyze 150~ks of archival \textit{Chandra} data to constrain the hot phase properties of NGC~1266 as well as derive warm gas electron number densities and temperatures from archival MUSE data. We summarize our results as follows:

\begin{itemize}
    \item We find temperatures between \edit1{0.25 and 1.8} keV peaking near the center of NGC~1266. We derive high electron number densities uncommon in starburst and AGN-driven galactic winds, ranging from $0.33-4.2\;\rm{cm^{-3}}$. Assuming a filling factor of one, these densities imply high mass outflow rates and short cooling times of order 1~Myr, indicative of the outflows undergoing rapid radiative cooling. The derived outflow rates are in excess of the cold molecular constraints from the literature, and this discrepancy can be resolved if the filling factor is less than unity, but that would lead to even greater electron densities.
    \item When considering the spectra from the whole northern and southern outflows, we find that charge-exchange (CX) statistically improves the spectral fitting of the southern lobe but not of the northern lobe. The CX emission makes up over a third of the total broad-band X-ray emission in the southern outflow, implying abundant interactions between the hot phase and neutral ISM. Along with M51 \citep{Yang2020}, NGC~1266 is one of the few AGN systems were CX has been detected to date.
    \item We constrain the thermal energy of the outflow to be in the range of $10^{54}-10^{56}$~erg. This is in excess of the previous minimum jet energy of $1.7\times10^{54}$~erg \citep{Nyland2013}. Thus, regardless of whether the radio jet solely drives the outflow, the hot phase composes a large fraction of the energy budget.
    \item We reprocess the MUSE data for NGC~1266 using the PHANGS-MUSE pipeline \citep{Emsellem2022}. We find a cavity-like feature in the southern outflow most evident in the H$\beta$ map. The ratio of H$\alpha$/H$\beta$ is in excess of 10 in the cavity, implying heavy extinction. The \ion{S}{2} doublet ratio also reaches the low-density limit in the cavity, implying very diffuse material is present. Based on the kinematic map of \citep{Eskenasy2024}, we hypothesize that the cavity is the receding side of the outflow; however a full kinematic analysis in future work (Otter et al. 2026 in prep) is necessary to determine its origin. 
    \item We find spatial agreement between the X-ray and H$\alpha$ images, the latter of which has been shown to originate from shocks. Thus it is likely the hot wind is shocking the ambient medium and producing the H$\alpha$. We find that the X-ray emitting gas is over-pressurized compared to the warm gas in the inner wind, but the two phases approach pressure equilibrium at the maximum wind extent of $\sim$750~pc indicating the wind may stall there barring future AGN activity. 
    \item We compare our results to other nearby AGN and starburst systems. We find that while the thermal energies are similar between NGC~1266 and other AGN, NGC~1266's hot wind is denser by a factor of a few. When comparing with starbursts, we find NGC~1266 is also denser though only near the AGN and is similar in density past $325$~pc. 

\end{itemize}

\software{CIAO (v4.15; \citealt{CIAO2006}), XSPEC (v12.13.0c; \citealt{XSPEC})}

\begin{acknowledgements}

SL and LAL were supported by NASA’s Astrophysics Data Analysis Program under grant No. 80NSSC22K0496SL, and LAL also acknowledges support through the Heising-Simons Foundation grant 2022-3533. SL and LAL also thank the OSU Galaxy/ISM Meeting for useful discussions and Tim Davis for helpful feedback on the optical line data analyses. 
\edit1{We also acknowledge the usage of the HyperLeda database \hyperlink{http://leda.univ-lyon1.fr}{http://leda.univ-lyon1.fr}} 

\end{acknowledgements}

\bibliographystyle{aasjournal}
\bibliography{main}{}

\end{document}